\documentclass[twocolumn]{emulateapj}
\slugcomment{To appear in ApJ}

\shorttitle{Hemispheric Pattern in Current Helicity }
\shortauthors{Gosain et al.}

\begin{document}
%\title{Hemispheric Pattern of Current Helicity Density from Synoptic Maps of Photospheric Vector Magnetic Fields }
\title{First Synoptic Maps of Photospheric Vector Magnetic Field from SOLIS/VSM:\\ Non-Radial Magnetic Fields and Hemispheric Pattern of Helicity}

\author{S. Gosain\altaffilmark{1} A. A. Pevtsov\altaffilmark{1} G. V. Rudenko\altaffilmark{2} and
S. A. Anfinogentov\altaffilmark{2}}

\altaffiltext{1}{National Solar Observatory, 950 N Cherry Avenue, Tucson 85719, Arizona, USA}
\altaffiltext{2}{Institute of Solar-Terrestrial Physics (ISTP),
Russian Academy of Sciences, Irkutsk, Russia}

\begin{abstract}
We use daily full-disk vector magnetograms from
Vector Spectromagnetograph (VSM) on Synoptic Optical Long-term Investigations
of the Sun (SOLIS) system to synthesize the first Carrington maps of the
photospheric vector magnetic field. We describe these maps and  make a comparison of observed radial field with the radial field estimate from LOS magnetograms. Further, we employ these maps to study the
hemispheric pattern of current helicity density, $H_c$, during the
rising phase of the solar cycle 24. Longitudinal average over the
23 consecutive solar rotations
shows a clear signature of the hemispheric helicity rule, i.e.
$H_c$ is predominantly negative in the North and positive in South.
%The hemispheric pattern for individual Carrington rotations is
%statistically weak, consistent with previous studies of
%active regions' helicity.
Although our data include the early phase of cycle 24, there appears
no evidence for a possible (systematic) reversal of
the hemispheric helicity rule at the beginning of cycle as
predicted by some dynamo models.
Further, we compute the hemispheric
pattern in active region latitudes (--30$^\circ \le \theta \le$ 30$^\circ$) separately for
weak (100$< |B_r| <$500 G)and strong ($|B_r|>$1000 G) radial magnetic fields. We
find that while the current helicity of strong fields follows the
well-known hemispheric rule (i.e., $\theta \cdot$H$_c <$ 0), $H_c$ of weak fields exhibits an
inverse hemispheric behavior (i.e., $\theta \cdot$H$_c >$ 0) albeit with large statistical scatter.  We discuss two plausible scenarios to explain the opposite hemispheric trend of helicity in weak and strong field region.
\end{abstract}

\section{Introduction}
Solar magnetic fields exhibit a hemispheric preference in their sense
of twist or helicity. Using chromospheric H$_\alpha$ images
\cite{hale1927} studied super-penumbral whirls around sunspots. He found that similar to terrestrial hurricanes, sunspot whirls exhibit
hemispheric preference in their shape. Later,
\cite{richardson1941} verified results of \cite{hale1927} by studying a
larger data set and found a hemispheric preference at $\sim$70\% level, although
only about one-third of the sunspots showed H$_\alpha$ vortices.
Origin of twist in super-penumbral whirls, initially believed to be due to
Coriolis force acting on plasma flows, is now attributed to presence of
electric currents in sunspot magnetic fields. Later studies establish what is now
known as the hemispheric helicity rule in various solar features associated with
magnetic fields: chromospheric filaments
\citep{rust1999,martin2008}, super-penumbral whirls
\citep{bala2004}, sheared coronal arcades
\citep{rust1996,canfield1997}, and interplanetary magnetic field
\citep{smith1999}.

Observations from modern vector magnetographs enabled researchers
to carry out a quantitative study of magnetic/current helicity and its sign
(chirality) \citep{seehafer1990,pevtsov1995,abramenko1996}. The chirality from
magnetograms is deduced by deriving the quantity,
$\alpha=(\nabla\times\vec{B})_z/B_z=J_z/B_z$, where $J_z$ is the
vertical component of electric current density. This parameter $\alpha$
is also known as the force-free parameter following the definition of
force-free fields for which the Lorentz force is zero, i.e.,
$\vec{J}\times \vec{B}=0$, which in turn implies $\vec{J}=\nabla \times
\vec{B}= \alpha \vec{B}$. Although the gas pressure and magnetic
pressure are comparable at photosphere and so photospheric fields are
not completely force-free \citep{priest1984,metcalf1995}, such approximations are
widely used to extrapolate magnetic fields in the corona using
photospheric magnetic field measurements.  The observations show a
hemispheric preference for  the sign of $\alpha$ parameter in solar
active regions with preferentially  negative values in the North and
positive values in the Southern hemisphere
\citep{seehafer1990,pevtsov1995,longcope1998}.
%However, statistically
%the hemispheric preference deduced from magnetograms is much weaker and
%has lot of scatter, unlike the hemispheric preference deduced
%morphologically.
Similar results are obtained by analyzing the vertical
component of current helicity density $H_c^z$, or simply $H_c$, given by $H_c=J_z\cdot B_z$, whose sign
measures the sense of twist of the magnetic field
\citep{abramenko1996,bao1998}. Here one must bear in mind that only vertical component of current helicity density is measured from vector magnetograms observed at a single height, the other two components can in principle change the sign of true current helicity density. However, under the assumption that the sense of flow (parallel or antiparallel) of the vertical component of electric current, $J_z$, with respect to $B_z$, is same as that of current vector $\vec{J}$ along $\vec{B}$, we can treat the vertical component of current helicity density as a measure of twist.
\cite{seehafer1990} showed that for cylindrically symmetric flux tube magnetic and current helicity have same sign and increase with each other. Further, it can be shown that $\alpha$ and $H_c$ are related
to each other: $H_c = \alpha B_z^2$ \citep{Hagyard1999}.

The hemispheric helicity rule is also present in large scale magnetic
fields (LSMF) \citep{pevtsov2000,wang2010}. In full sun MHD simulations,
\cite{yeates2008} found similar hemispheric preference but only at
mid/low latitudes. At higher latitudes they found a reversal of the
hemispheric pattern of twist, which is in contrast to the observations
which show that polar crown filaments having chirality preference of
the same nature as active regions. In addition to the hemispheric helicity rule,
the helicity of LSMF was found to show evidence of zonal organized bands
\citep{pevtsov2003} co-spatial with patterns of torsional oscillations
\citep{howe2000}. Sign of helicity in these bands is opposite to $H_c$ prevailing
for a given hemisphere.

Until recently, the full disk vector
magnetograms were not routinely available, and
thus, early studies of $H_c$ in
large-scale fields were based on pseudo-vector derivations.
In this method, the components of LSMF
vector are derived from time sequence of longitudinal
magnetograms under the simple assumption that the field
does not change over several days. Even with this assumption, the
derivations are limited to only two components (radial and toroidal),
the variation in projection are too small to derive the meridional component
of the field (however, see \cite{wang2010}). The routine observations of full disk vector magnetic fields
by Vector Spectromagnetograph (VSM) on Synoptic Optical Long-term Investigations
of the Sun (SOLIS) system \citep{keller2003} and Helioseismic Magnetic Imager (HMI) \citep{schou2012} on
Solar Dynamic Observatory (SDO) spacecraft, allow for the first time to investigate the
current helicity of large-scale magnetic field directly without any restrictive
assumptions about the nature of these fields.

%Observationally we need to carry out a systematic study of the magnetic
%field vector in solar active regions over long timescales to establish
%spatio-temporal patterns in the twistedness of the magnetic field in
%solar active regions. The questions such as: whether there is
%systematic variation in helicity pattern with solar cycle?, Whether
%strong and weak fields show similar hemispheric pattern?, How does
%helicity pattern in active region belt affect polar fields?, and so on,
%all need systematic long-term observations of the vector magnetic
%fields of the sun. The subsurface origin of magnetic fields, their rise
%to the photosphere and subsequent conversion of toroidal field to
%global dipolar field is not understood very well and there are
%competing models of solar dynamo which try to explain the observed
%features of solar cycle.  The helicity patterns, which systematic
%studies of vector fields on the sun will establish further, is
%therefore another important property that any dynamo model should
%explain.  Observations with such goals are taken routinely with the
%Vector Spectro-Magnetograph (VSM) instrument of SOLIS (Synoptic Optical
%Long-term Investigations of the Sun) project of National Solar
%Observatory (NSO) \citep{keller2003}, which we use in this work.

In this paper we present the first ever synoptic Carrington maps of the
vector magnetic field constructed from VSM/SOLIS daily observations.
We contrast the new maps with traditional Carrington charts
derived from longitudinal field measurements and find significant
differences between true and LOS-based radial magnetic field in areas of
active regions and at high latitudes. In Section 2
we briefly describe the SOLIS/VSM instrument and the method of deriving the
full disk vector magnetograms. Then we describe synoptic maps of vector magnetic field and the properties of the radial magnetic field in Sections 3 and 4. In Section 5 we present the analysis of the current helicity density based on new synoptic maps. In Section 6 we study the helicity pattern in strong and weak magnetic field regions separately, and in Section 7 we discuss our findings.

\section{Observations and Data Analysis}

We employ daily observations of vector magnetic fields taken
with the Vector Spectromagnetograph (VSM)- one of three instruments comprising
SOLIS facility for synoptic observations of the Sun in
optical wavelengths \citep{keller2003,bala2011}.

%One of the
%instruments in SOLIS suite is  Vector Spectro-Magnetograph
%(VSM)\footnote{for more details see
%http://solis.nso.edu/VSMOverview.html}.
%SOLIS/VSM instrument routinely
%makes fulldisk magnetic field measurements of the Sun.

SOLIS/VSM is spectrograph-based spectropolarimeter. It takes the full Stokes profiles
of the Fe I 630.15 -- 630.25 nm line pair along
slightly curved spectrograph slit that intersects the entire solar disk (from East to West limbs).
Full disk magnetogram is built by scanning the solar image by moving the telescope in
the declination. Pixel size in final magnetogram is about 1$\times$1 arcsec$^2$.
VSM takes about 0.6 seconds to record all four Stokes parameters for
a single scan line, and it takes about 20 minutes
to complete a full disk magnetogram (2048 scan lines).

The Stokes I, Q, U, and V profiles observed in Fe I 630.15-630.25 nm line pair
are sampled in spectral direction
with 2.4 pm pixel$^{-1}$. The spectra are inverted in a framework of Milne-Eddington
model of stellar atmosphere following Unno-Rachkovsky formalism
%and local thermodynamic equilibrium (LTE) assumptions to infer the
%magnetic field vector over the full-disk
\citep{skumanich1987}.

Additional details about the instrument and the pipeline reduction steps
can be found elsewhere (e.g.,
%The data reduction pipeline for the processing of full-disk vector
%magnetograms is described in
\citep{jones2002,henney2006,bala2011}.
Only pixels with polarization signal above the threshold
of 0.1\% of continuum intensity, $I_c$, are inverted to obtain
the magnetic (field strength, inclination angle and
azimuth angle) and thermodynamic (e.g., Doppler width, Doppler velocity,
source function, temperature) parameters. The method also allows to determine
the relative contribution of magnetic and non-magnetic plasma to line profile
for each pixel (filling factor). The threshold of 0.1\% of $I_c$ corresponds to typical noise level in the continuum. Using this threshold avoids fitting profiles buried in the noise. The error in inferred magnetic field parameter for each pixel is different as Stokes signal varies from pixel to pixel.  For example, \cite{gosain2010} used {\it Hinode} vector magnetogram of a sunspot region as reference field and simulated random errors due to normally distributed photometric noise (0.5\% of $I_c$, a  3-$\sigma$ noise level) in Stokes profiles and found that maximum error in the magnetic field parameters is $\sim$ 50 G for field strength, $\sim$ 1.5$^\circ$ for inclination and $\sim5^\circ$ for azimuth angle, respectively. Stokes profiles in regions like sunspots, plage, network and decaying regions have typically SNR $\geq$1000 and so the typical errors are expected to be in this range. The noise levels in SOLIS magnetograms are estimated to be a few Gauss in the longitudinal and 70G in the transverse field measurements \citep{tadesse2013}. %However, one should bear in mind that systematic errors, which arise due to simple assumptions such as neglect of depth stratification in Milne-Eddington model, could be much larger than the random errors. }

The 180 degrees azimuth ambiguity is resolved using Non-Potential Field Computation (NPFC) method
\citep{manolis2005,manolis2008}. Due to limited memory and CPU speed constrains in early data reduction pipe-line
hardware, the 180-degree disambiguation is parallelized by dividing the solar image
into smaller overlapping tiles.  The azimuth ambiguity is resolved in each tile
independently. For this study, we have also tested a new faster ambiguity
resolution method developed by \cite{rudenko2011}.
In this method, the direction of the transverse field is determined in
accordance with the principle of minimum deviation of net differences of
the potential field from those of the defined field. The results presented in this paper are found to
be same with both ambiguity resolution algorithms, which tests the
validity and consistency of the new and faster algorithm developed by
\cite{rudenko2011}.

\section{Synoptic Maps of the Magnetic Field Vector}

To compute helicity of magnetic fields over the entire solar disk
requires constructing the synoptic maps of vector field. Until now,
no such maps were produced. Here we describe our approach to constructing
the vector field synoptic maps and compare these new maps with traditional
synoptic maps of radial field, which are created from LOS
magnetograms under a restrictive assumption
that magnetic field are normal to solar surface.

Traditionally, synoptic (or Carrington) maps are synthesized by combining the daily full-disk
magnetograms taken over the period of one solar rotation ($\approx$ 27 days).
The maps cover all latitudes ($\pm$ 90 degrees) and longitudes (0--360 degrees)
of solar surface.
Although the purpose of synoptic map is to represent activity over entire
surface of the Sun, the
maps are representative of the activity occurring near the central meridian  (CM) of
the solar disk at the time of observations.  In earlier times when the angular resolution of data
was coarse a cos$^{4}\phi$ weighting was used \citep{harvey1980}, where $\phi$ is the
longitudinal distance from the central meridian. With the availability
of higher-resolution data, it has been found that such weighting leads to
a smearing of small scale features which evolve substantially from day
to day. A Gaussian weighting scheme of the form
$W(\phi)=\exp^{-(\phi/7)^2}$ has therefore been implemented (J. W. Harvey,
private communication) in order to heavily weight only the regions very
close to the CM. Such synoptic maps, therefore, essentially capture a
snapshot of activity very close to the CM over the duration of solar
rotation.

Synoptic maps based on the LOS magnetic field of the Sun
are widely used to study the evolution of magnetic fields
on time scales from solar rotation to the solar cycle. The maps of radial
field, constructed from LOS magnetograms are used as the lower
boundary condition  for
extrapolating the photospheric fields into the corona in a framework
of potential field extrapolation \citep{altschuler1969}.
The current-free fields extrapolated to the heights in the corona where they deemed
to become purely radial (Potential Field Source Surface, or PFSS,
typically between 2.5 and 3.5 $R_{\odot}$) are used in modeling the coronal and
heliospheric magnetic fields, the solar wind speed and the appearance of white-light
corona prior to total solar eclipses. The validity of the assumption that the magnetic field is normal to the surface at the photosphere
level is largely unknown. Thus, the availability of synoptic maps of vector magnetic field allows us to verify the assumption of field verticality for the first time.

Below we describe the steps taken in constructing the vector synoptic maps.

\begin{itemize}
\item{After the 180-degree ambiguity is resolved,
the three components of the field in image plane ($B_x$ - (terrestrial)
East-West direction,
$B_y$ - (terrestrial) North-South component,
and $B_z$ -- vertical) are transformed to heliographic coordinate system which
yields toroidal ($B_\phi$), poloidal ($B_\theta$) and radial ($B_r$) component.
The full-disk magnetograms are then re-mapped into heliographic
coordinates (longitude: $\phi$, latitude: $\theta$).}

\item{The individual re-mapped magnetograms are combined in traditional way with
a Gaussian weight of the form $W(\phi)=\exp^{-(\phi/7)^2}$ given to
the data. This results in three synoptic maps (one for each component of vector field) corresponding to
one Carrington rotation.}

\end{itemize}

\begin{figure}
\epsscale{0.65}
\plotone{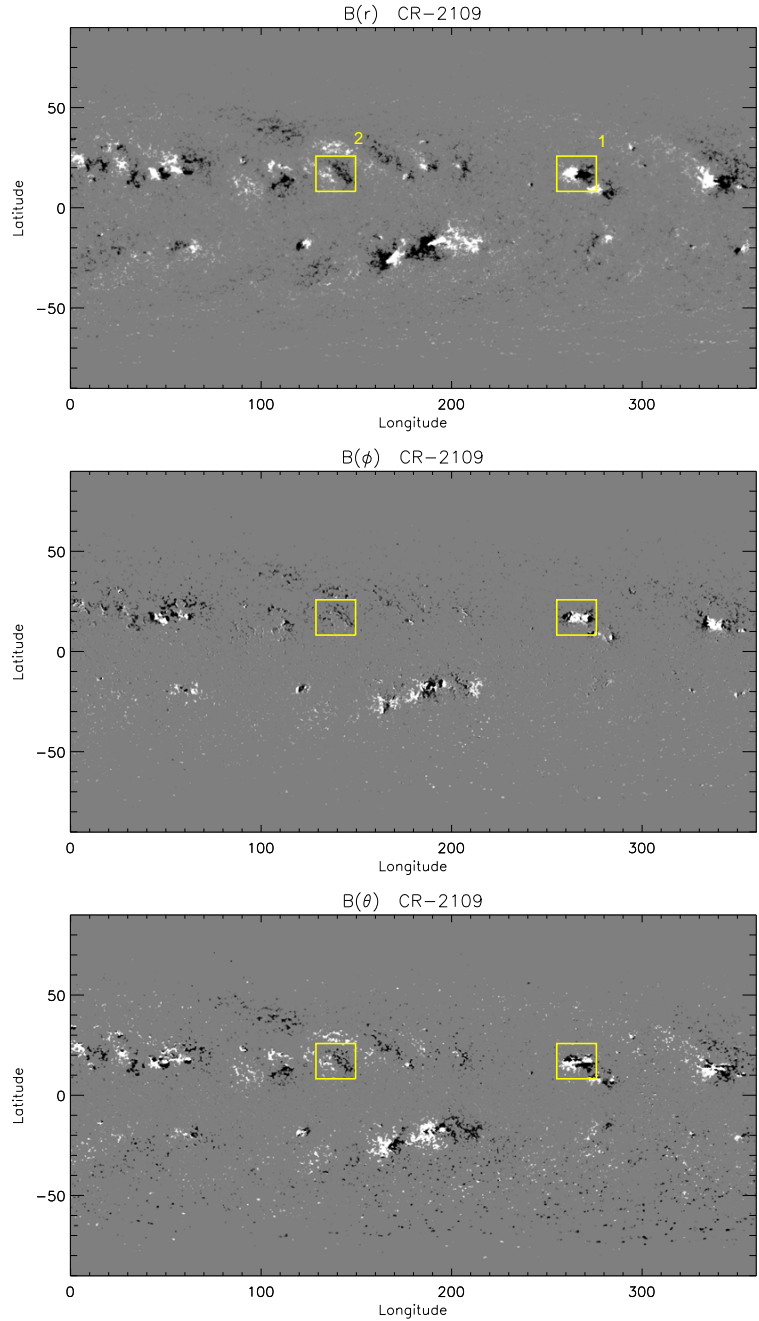}
\caption{Synoptic Carrington map of the vector magnetic field components synthesized using full disk SOLIS/VSM vector magnetograms is shown for CR-2109. The panels from top to bottom show the distribution of $B(r), B(\theta)$ and $B(\phi)$ components, respectively. The $B_r$ map is scaled between $\pm$ 100 G and $B_\theta$ and $B_\phi$ maps are scaled to $\pm$ 20 G. The positive values of $B_r$, $B_\theta$ and $B_\phi$ points respectively, upwards, Southwards and to the right (Westward). The zoom-in of the regions marked by rectangles `1' and `2' is shown in Figure~\ref{vector_roi}.}
\label{synoptic}
\end{figure}

\begin{figure}
\epsscale{0.71}
\plotone{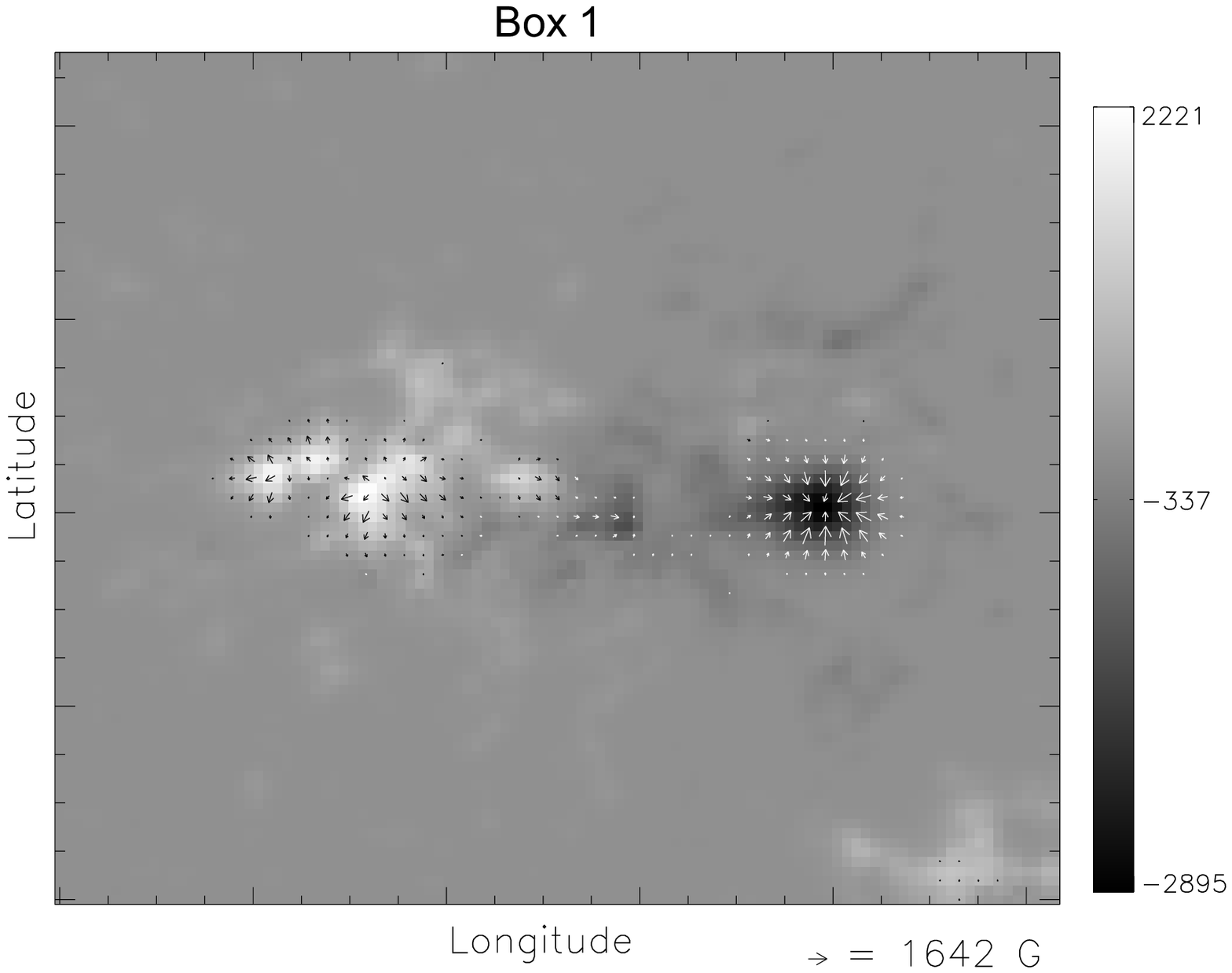}
\plotone{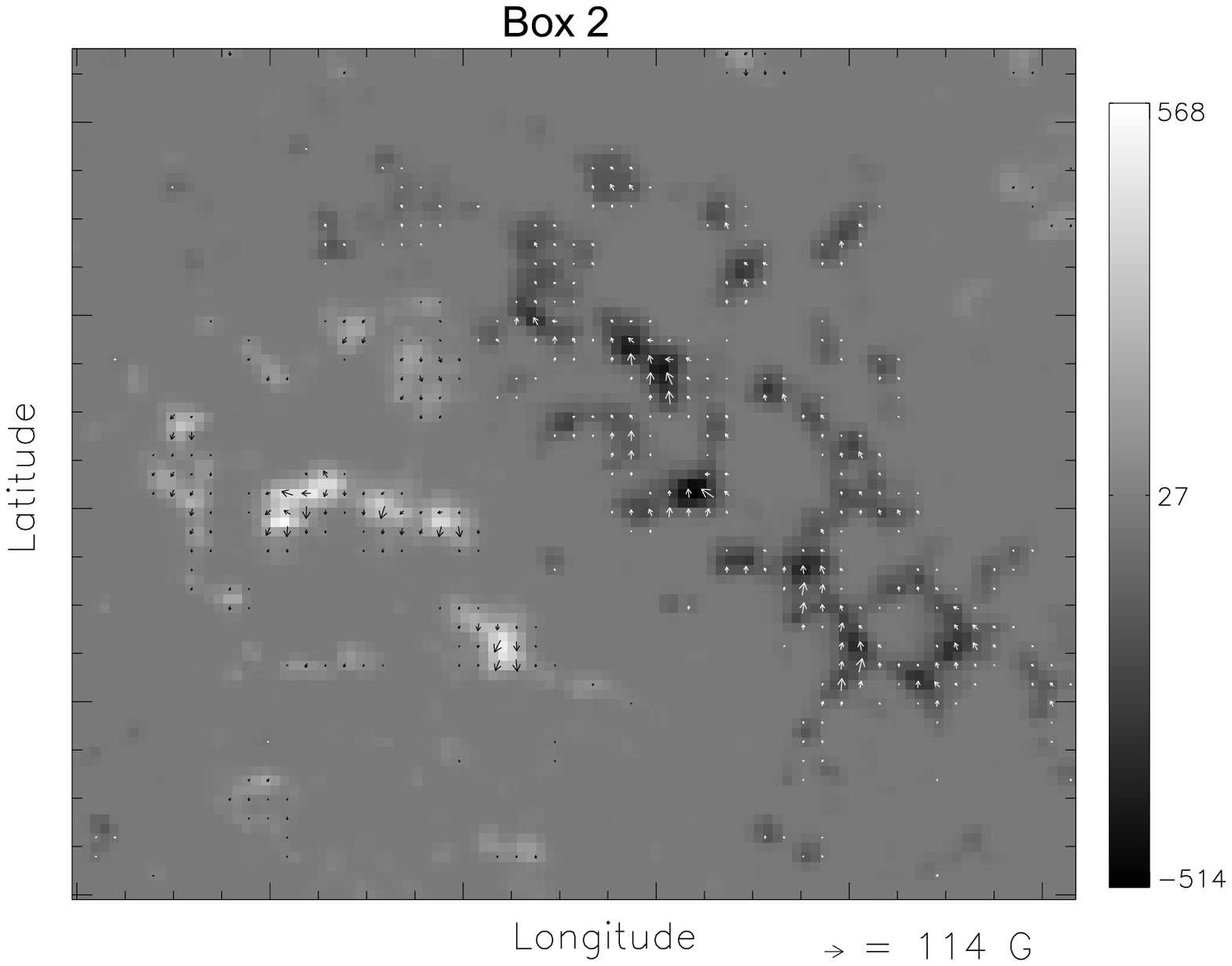}
\caption{The zoom-in of the two regions marked by rectangles, `1' and `2', in Figure~\ref{synoptic} is shown. The arrow at the bottom and color bar on the side of the panels show scale of transverse and radial field, respectively. The tick-marks on the latitude and longitude axis represent 1 degree increments.   }
\label{vector_roi}
\end{figure}

As an example, in Figure~\ref{synoptic} we show a vector synoptic map for  Carrington rotation 2109.  The zoom-in of a  typical active region (Box `1') and a diffuse bipolar region (Box `2') along with the overlaid transverse vector is shown in the Figure~\ref{vector_roi}. We find the active region field pattern to be consistent with expectation for dipolar field configuration: $B_\phi$ component (Box`1') between two polarities corresponds to field lines connecting them, and $B_\theta$ is in agreement with the group tilt corresponding to Joy's law \citep{hale1919}.  The patterns for large scale field (Box `2') are similar to those obtained by \cite{pevtsov2000} with the pseudo-vector method  for their data during cycle 22. Both foot points of the field in Box `2' show negative $B_{\phi}$, which means that the field lines are connected such that the foot points make obtuse angle with solar surface, measured from polarity inversion  line (PIL). Further, the transverse field vectors in diffuse bipolar region (Box `2') show a North-South component ($B_\theta$) such that the field configuration is tilted towards equator. The reason for the equator-ward tilt could be  presence of a filament along the PIL would suggest a non-potential field oriented along the PIL, which is oriented roughly along N-S direction. Further, the period of this observation (29-April-2011)  corresponds to ascending phase of solar cycle 24, so coronal equatorial streamers which are prominent during solar minima, may also give the large scale field (reaching high in the corona), a net equator-ward tilt. Thus, the derived components of vector magnetic field in our synoptic maps agree with the expected orientation of magnetic fields on the Sun.

\section{Observed Radial Field versus Radial Field Estimate from Longitudinal Magnetograms}

Assuming that the magnetic field on the Sun is mostly vertical, one can estimate the radial field, $B_{r(LOS)}$, using the relation $B_{r(LOS)}=B_{LOS}/\mu$, where $\mu$ is the cosine of the heliocentric angle and $B_{LOS}$ is the line-of-sight field.  Traditional synoptic maps of radial field are synthesized using this method, which will work exactly if the field were truly normal to the solar surface. However, in practice magnetic fields are not normal everywhere. For example, the fields are more horizontal in sunspot penumbra and near PIL. Although, in most cases the field in quiet Sun appears to be more vertical, it is not purely normal to the solar surface \citep{gosain2012}. The departure of high latitude fields from radial expansion at the photospheric level was deduced using vector field reconstructed from  longitudinal magnetograms by \cite{petrie2009}.

Now let us compare the radial field reconstructed from longitudinal measurements, $B_{r(LOS)}$, with the observed radial field component from vector field measurements, B$_{r(OBS)}$ (or simply $B_r$, for brevity). For comparison we compute the difference of the absolute  value of two radial fields, i.e., $\Delta B_r = |B_{r}|-|B_{r(LOS)}|$,  and plot signed relative difference, $\Delta B_r / |B_r|$ (\%), as shown in the top panel of Figure~\ref{diffmaps}. This signed difference map is saturated at $\pm$5\% to emphasize the latitudinal pattern in the sign of the difference. Black (white) correspond to negative (positive) difference, i.e., $|B_r| < |B_{r(LOS)}|$ and $|B_r| > |B_{r(LOS)}|$, respectively.  It can be noticed that in active regions the differences are large as compared to fields outside active regions at that latitudes. This is expected due to the presence of fanning fields in active regions.

Further, in the middle panel of Figure~\ref{diffmaps} we show the longitudinal average of absolute value of relative difference, $|\Delta B_r| / |B_r|$ (in \%). One can notice that the  value of  relative difference increases systematically towards higher latitudes. The observations are scanty at higher latitudes, hence a large scatter is seen, but the systematic increase towards higher latitudes is evident from the profile.
To understand the cause of this systematic increase let us consider a field of strength $B$ at latitude $\theta$  having an inclination $\pm\gamma$ in the meridional plane with respect to local vertical. The positive or negative sign corresponds to case when the field is inclined towards poles or solar equator. Then the LOS field, $B_L$ and radial field, $B_r$ would be given by
$$B_L=B \cdot cos(\theta\pm\gamma)$$
$$B_r=B \cdot cos(\gamma)$$

When the assumption is made that the field is normal to the solar surface ($\gamma=0$), $B_r(LOS)$ is derived as $B_L/cos(\theta)=B cos(\theta \pm 0)/cos(\theta)$. But if the field is not vertical ($\gamma \ne 0$), this assumption introduces an error. In the latter case, $B_r(LOS)=B cos(\theta \pm \gamma)/cos(\theta)$.

Difference ($\Delta B_r$) between true $B_r$ and one derived from assumption that field is vertical can be expressed as following:

$$\frac{\Delta B_r}{B_r}= 1-\frac{cos(\theta \pm \gamma)}{cos(\theta)cos(\gamma)} $$

One can show that
$$\frac{\Delta B_r}{B_r}=\frac{\Delta B_r}{|B_r|}= \pm tan(\theta)tan(\gamma)$$

The expression for normalized amplitudes of difference between true $B_r$ and $B_r(LOS)$ becomes

$$\frac{|\Delta B_r|}{|B_r|}=tan(\theta)tan(\gamma)$$

Thus, even if $\gamma$= constant, $|\Delta B_r| / |B_r|$ will vary as tangent of latitude ($\theta$). The profile in middle panel of Figure~\ref{diffmaps}  demonstrates that the effects of non-verticality of magnetic field become stronger (and more important) for high latitudes even if the inclination angle does not change systematically.

%For comparison we compute the difference of the absolute value of two radial fields, i.e.,
%Br = |Br(OBS)| - |Br(LOS)|, as shown in the top
%panel of Figure 3. The map is scaled to ± 1 Gauss,
%to highlight the sign of the difference value. The
%negative (black) and positive (white) values of
%deltaBr mean that Br(LOS)
%is stronger and weaker,
%respectively, as compared to Br(OBS)
Further, it can be noticed that there is a systematic pattern
in the sign of the relative difference, $\Delta B_r / |B_r|$, as described
below.
\begin{itemize}
\item{Both the active regions as well as the diffuse field outside active regions show asymmetry in the sign of $\Delta B_r / |B_r|$ along the North-South direction.}
\item{In both hemispheres, the portion of the active region towards the equator (pole) shows negative (positive) values of $\Delta B_r / |B_r|$. }
\item{Outside active regions the diffuse flux in both hemispheres shows opposite pattern as compared to active regions, i.e., the flux near the equator (pole) shows positive (negative) value of $\Delta B_r / |B_r|$.}
\end{itemize}

%
%In order to understand this pattern of $\Delta B_r$, first let us consider a purely vertical field, $\vec{B}=B\cdot\hat{n}$, of uniform magnitude, distributed at each latitude on the sun. Then by definition
%$\Delta B_r=0$ at all latitudes. Now let us introduce a systematic tilt of 5$^\circ$ in the field $\vec{B}$ towards the equator in both hemispheres of the sun. For such a case the normalized value of the difference, $\Delta B_r / |B_r|$, is shown by dash-dotted curve in the middle panel of Figure~\ref{diffmaps}. The difference will be negative in sign and the relative error would grow to infinity towards the poles. Similarly, if we now consider the case when a systematic tilt of 5$^\circ$ is introduced in the field, $\vec{B}$, towards the poles in both hemispheres of the sun. The difference in this case would be positive in sign and relative error would again grow to infinity near the poles.
%
%Now let us understand the observed pattern of $\Delta B_r$ as seen in top panel of Figure~\ref{diffmaps}, by a simple picture.
Negative $\Delta B_r / |B_r|$ in high latitudes in Northern and Southern hemispheres implies a systematic tilt of vector magnetic fields towards equator.
The bottom panel of Figure~\ref{diffmaps} illustrates a cartoon of the side view of portion of the solar disk visible to observer (labeled LOS) and some example field lines (dashed curved lines). The labels correspond respectively as follows: North Pole:NP, South Pole:SP, Equator: EQ, Radial Direction: R, Center of Sun: O, Active Region: AR, Low Latitude: LL, and High Latitude: HL.
It can be seen that in active region (labeled AR), depicted like a sunspot with the field lines fanning out, the equator-ward portion of AR will have field lines oriented in such a away that they deviate from local solar vertical, inclining towards the equator. Similarly the pole-ward portion of AR will have field lines oriented in such a away that they deviate from local solar vertical, inclining towards poles. Then  the field inclined towards equator (pole) will show negative (positive) sign of $\Delta B_r / |B_r|$, which is what we observe in top panel of Figure~\ref{diffmaps}.
 On the other hand the large scale field lines (depicted by dashed curved lines) rooted in HL and LL deviate from local solar vertical inclining equator-ward and pole-ward, respectively. Thus, we will see $\Delta B_r / |B_r|$ to be negative in HL regions and positive in LL regions, which is what we observe in top panel of Figure~\ref{diffmaps}.

These results emphasize importance of vector field measurements. The quantitative effects arising from replacing the B$_{r(LOS)}$ synoptic maps by observed  B$_{r}$ on solar wind and coronal field extrapolation are unknown. These effects and detailed study of non-radial nature of fields will be a subject to our separate future study. Here we want to emphasize the importance of vector field measurements and the synoptic maps in deriving the $B_r$, and potentially calibrate the systematic deviation $\Delta B_r / |B_r|$ with latitude.

 The systematic pattern shown in top panel of Figure~\ref{diffmaps} cannot be a result of random noise. In Appendix-A we show that the random noise in $B_r$ and $B_r(LOS)$ will lead to random noise in relative difference and not a systematic variation in its sign as shown in top panel of Figure~\ref{diffmaps}.

\section{Current Helicity Density from Synoptic Vector Maps}

\begin{figure}
\epsscale{0.7}
\plotone{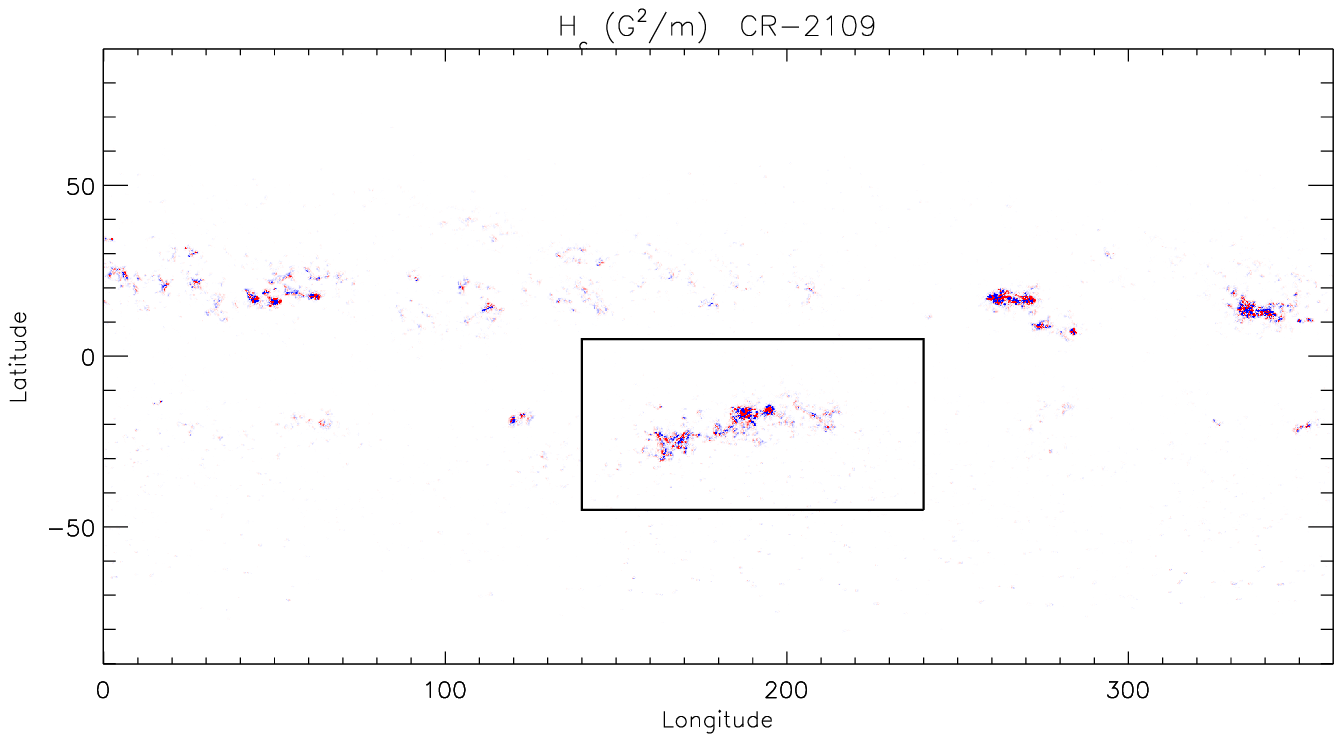}
\plotone{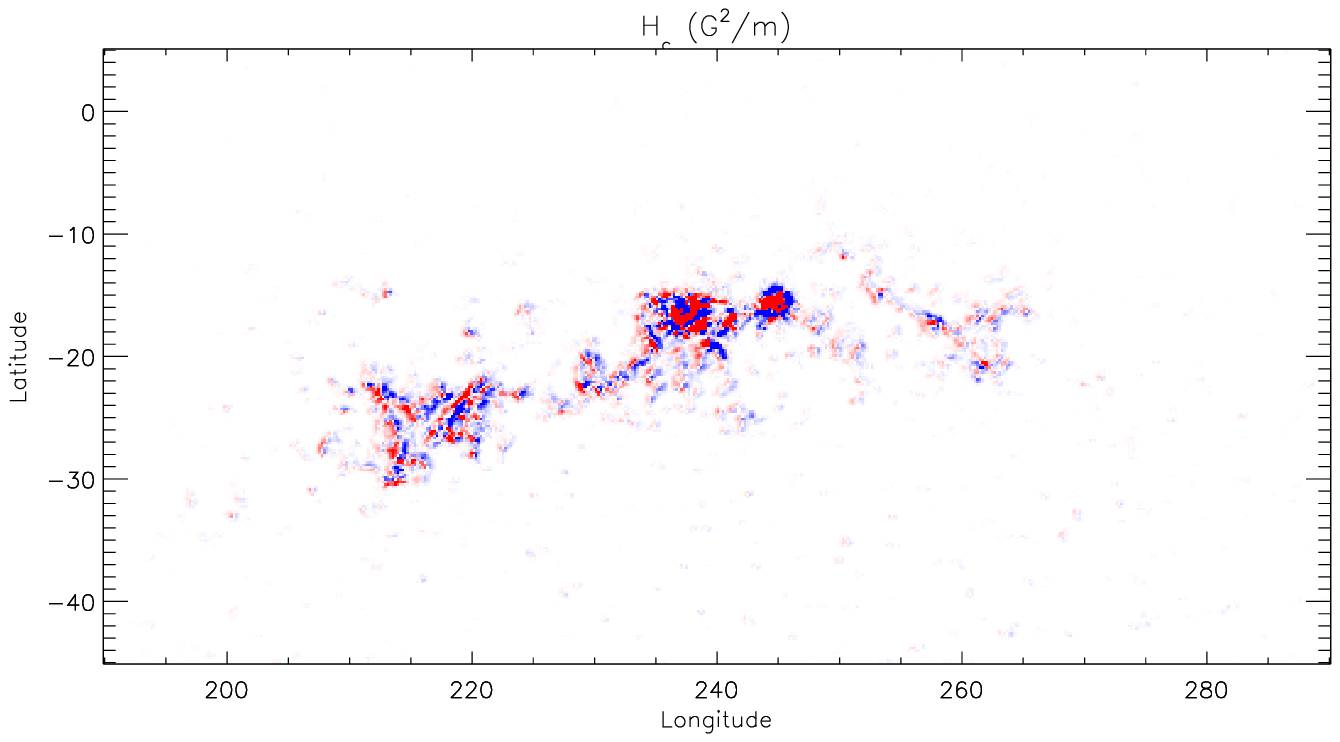}
\caption{Synoptic map of current helicity density, $H_c$, is shown in the top panel for CR 2109. The rectangular region marked in the map is zoomed and displayed in the lower panel. The amplitude of $H_c$ in the displayed images is scaled between $\pm2\times10^{-3}$ G$^2$ m$^{-1}$. }
\label{fig_hcmap}
\end{figure}

Given the distribution of vector magnetic field in the photosphere one can
compute the vertical component of current helicity density, $H_c$. In
spherical coordinates $$H_c(\phi,\theta)=B_r(\nabla\times\vec{B})_r$$ $$ =
\frac{1} {sin\theta} \{ \frac{\partial}{\partial\theta}[sin\theta B_{\phi}
(\phi,\theta)] - \frac{\partial B_{\theta} (\phi,\theta)}{\partial\phi} \}
B_r(\phi,\theta)$$ The distribution of $H_c$ for Carrington rotation (CR)
2109 is shown in Figure~\ref{fig_hcmap}. The patterns of current helicity
density of mixed sign can be seen in the active region. Such local helicity patterns
are well known from previous works \citep[e.g.][and references therein]{pevtsov1990,pevtsov1994,abramenko1996,pevtsov1998,su2009}.

%TBD starting from here

\begin{figure}
\epsscale{0.7}
\plotone{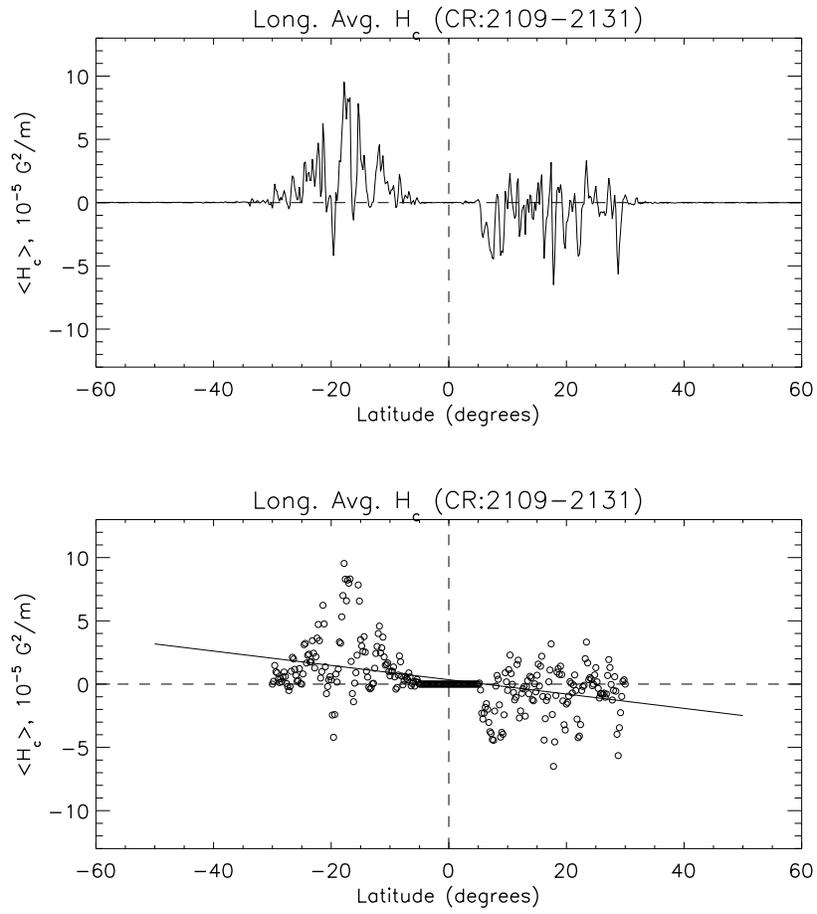}
\caption{The top panel shows longitudinal averaged profile  of $H_c$ over all CRs (2109 to 2131). The bottom panel shows the same profile between 0-30$^\circ$ along with a linear fit with a negative slope.}
\label{fig_linfit}
\end{figure}

In the present work we computed maps of vertical component of current
helicity density for CR 2109 to 2131 covering a period from March 2011 to
December 2012. During this time a total of 453 NOAA numbered active regions crossed
the solar disk. The longitudinal average of $H_c$, or $<H_c> $, distribution
over all the CR is shown in top panel of Figure~\ref{fig_linfit}. The $<H_c>$
profile shows a tendency to be positive in Southern and negative in Northern
hemisphere of the Sun, in agreement with the previous results obtained by
many researchers for active regions observed in cycles 22, 23 and even 24 \citep{abramenko1996,bao2000,pevtsov2001,hao2011}.
%However, we
%notice that the most of studies carried out earlier have combined
%observations of several individual active regions (AR) over limited field-of-
%view (FOV) and obtained such hemispheric pattern. In many cases a best fit
%($\alpha_{best}$) or spatially averaged value ($\alpha_{av}$) of twist
%parameter $\alpha$ is used for each AR. In the Figure~\ref{fig_linfit} we
%have all the longitudes averaged and therefore samples all of the flux seen
%near central meridian of the Sun during its disk passage and therefore  the
%distribution of $H_c$ we obtain is more complete as it covers a large range
%of magnetic fields on the sun.

In Figure~\ref{fig_linfit}, the $H_c$ is
averaged over regular ARs, ephemeral regions, large
scale bipolar and unipolar magnetic regions and plages up to high latitudes
($\sim$60$^\circ$). In the lower panel of Figure~\ref{fig_linfit} we show the
data points confined to active region belt (0-30$^\circ$) and we fit a
straight line to the observed data points to show that the slope of $d
H_c/d\theta$ is negative, indicative of hemispheric sign preference. In
computing the average profiles of $H_c$ we used only the pixels above the
threshold of 20 G for both radial and transverse fields to filter out the
noise.

\begin{figure}
\epsscale{0.8}
\plotone{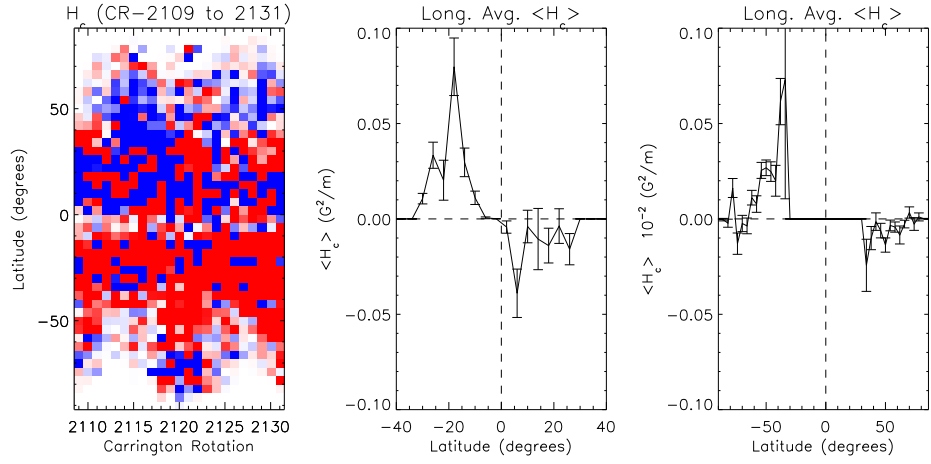}
\caption{Left panel shows the time-latitude plot of current helicity density, $H_c$. Each column corresponds to the longitude average of $H_c$ for the Carrington rotation and re-binned into 4$^\circ$ latitude bins. The blue (red) color represents the negative (positive) sign of $H_c$. The magnitude of $H_c$ is scaled between $\pm2\times10^{-4}$ G$^2 m^{-1}$. The middle and right panels show the mean latitudinal profile of $H_c$ (derived from time-latitude plot on left) over 0-30$^\circ$ and 30-90$^\circ$, respectively. The error bars show the standard error of the average $H_c$ value.}
\label{fig_timlat2}
\end{figure}

In the left panel of Figure~\ref{fig_timlat2} we construct the time-latitude
map of the $<H_c>$ during CR 2109-2131. Each column in the map represents the
longitudinal average of $H_c$, at all latitudes for the indicated CR number.
The blue and red colors represent the negative and positive sign of the $H_c$
where the color scale is saturated between $\pm2\times10^{-4} G^2 m^{-1}$.
The hemispheric pattern of $<H_c>$, can be seen visually by a dominance  of
negative (blue) in the North and positive (red) in the South hemisphere. The
patches of opposite sign are also present in each hemisphere consistent with
previous results that hemispheric rule is weak tendency.
Nevertheless on average we find that the hemispheric rule is followed during
the studied period which corresponds to the rising phase of current solar
cycle 24.

For large scale fields, several researchers have used pseudo-vector
reconstruction method \citep{pevtsov2000,wang2010} to compute $H_c$ and
reported that the hemispheric rule is followed by large scale fields as well. Here
in middle and right panels of Figure~\ref{fig_timlat2} we plot, respectively,
the average latitudinal profile (averaged over CR 2109 to 2131) in
0-30$^\circ$ and $>$30$^\circ$ latitude bands. The profile of $H_c$ in high latitude band ($>$30$^\circ$) is shown separately due to its relatively low amplitude. The error bars show the
standard error of the mean. These profiles show that hemispheric pattern of
$<H_c>$, namely negative in North and positive in South, is followed in both
latitude ranges quite well. Since the latitude range 30-90$^\circ$ contains
mostly large scale diffuse fields, we therefore confirm the results of
\cite{pevtsov2000} and \cite{wang2010} with observed vector magnetograms for
the first time. Another feature than can be noticed from these plots is that
the helicity pattern in southern hemisphere is much stronger in the sense that
we see fewer patches of opposite signed helicity there, as compared to the
Northern hemisphere where we see a weaker dominance of negative helicity. It
is well known that the North and South hemispheres show an asymmetry in the
amplitude and phase of their activity cycle. Here we show an evidence for
asymmetry in the strength of hemispheric dominance of the helicity sign.
Further, one can notice signatures of annual B-angle variation in the time-
latitude plot in Figure~\ref{fig_timlat2}. Such variation allows one to
sample vector magnetic field at higher latitudes and therefore, study the hemispheric
pattern there. We do see fluctuations in the sign of $<H_c>$ at high
latitudes, however, the average behavior is in agreement with general
hemispheric rule \citep{pevtsov1995}.

\section{Hemispheric pattern for the Weak and Strong Fields}
\begin{figure}
\epsscale{0.7}
\plotone{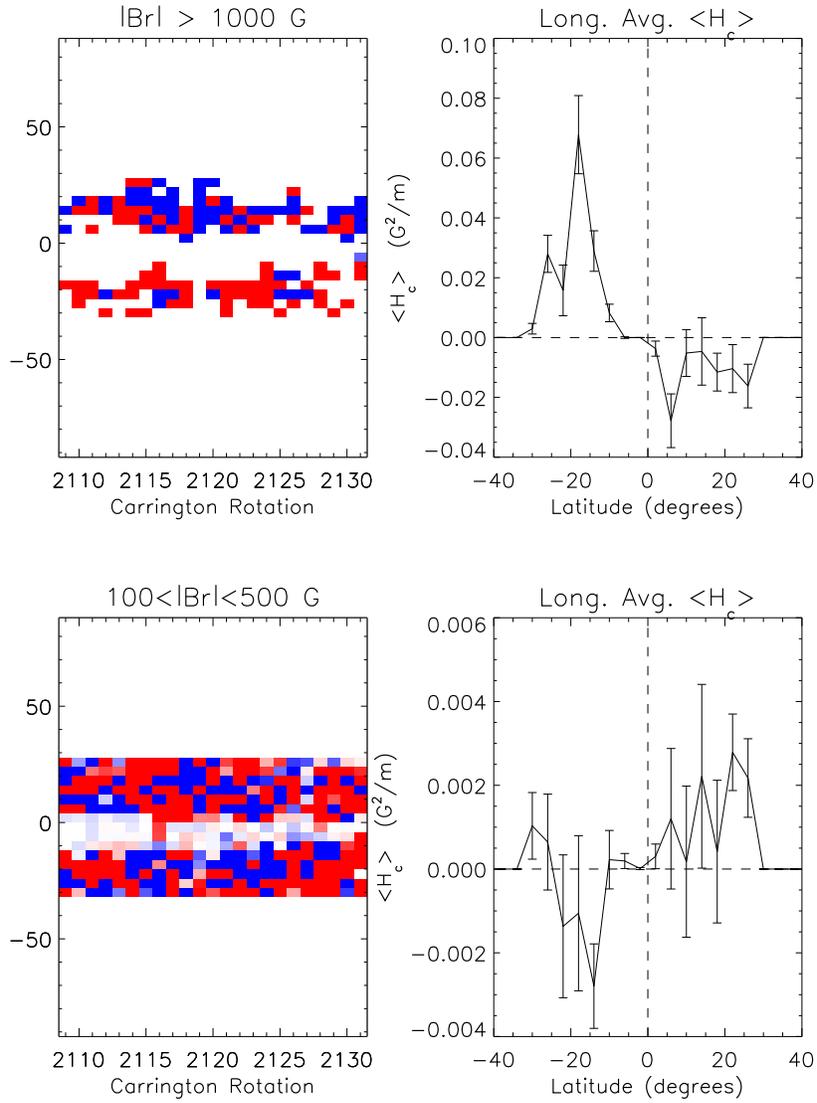}
\caption{Left top (bottom)  panels show the time-latitude plot of current helicity density, $H_c$ for strong (weak) fields over 0-30$^\circ$ latitude belt. Each column corresponds to the longitude average of $H_c$ for the Carrington rotation as labeled. The blue (red) color represents the negative (positive) sign of $H_c$. The magnitude of $H_c$ is scaled between $\pm5\times10^{-4}$ G$^2 m^{-1}$. The right panels (both top and bottom row)  show the mean latitudinal profile of $H_c$, derived from respective time-latitude plots on the left. The error bars show the standard error of the average $H_c$ value.}
\label{fig_strweak}
\end{figure}

Next, we study
separately the hemispheric pattern for strong and weak fields in the active
region belt 0-30$^\circ$. For segregating strong and weak field regions we
used the following criterion,{\it Strong Fields:} $|B_r| > 1000$ G and {\it
Weak Fields:} $100<|B_r|<500$ G. Such criterion was used by
\cite{mzhang2006}  in their study of hemispheric
pattern of helicity in active regions. Thus, adopting same criterion facilitates a straightforward comparison with their study. The top and bottom panels in
Figure~\ref{fig_strweak} show the results for strong and weak fields,
respectively. The panels on the left show the time-latitude plot of $<H_c>$
for active region belt between 0-30$^\circ$. Beyond 30$^\circ$ latitude only the fields weaker than 500 G remain, so we exclude these regions from comparison.
The panels on the right show the
latitudinal profile averaged over all 23 CRs (2109-2131) shown on the left
panels. It is found that the latitudinal profile of $H_c$ for the strong
field regions follows the hemispheric tendency and the pattern is similar to
one shown for 0-30$^\circ$ belt in Figure~\ref{fig_timlat2}, for all field
strengths. On the other hand for weak fields the latitudinal profile shows a
weak but systematic anti-hemispheric rule, i.e., preference for positive sign
of $H_c$ in North and negative in South hemisphere. Thus, our observations
which show that the strong fields obey hemispheric rule while weak fields
show tendency for anti-hemispheric rule, is in disagreement with the results
of  \cite{mzhang2006}, who found that weak fields obey the hemispheric rule while
strong fields follow anti-hemispheric rule. In another related study by
\cite{pevtsov2001b} it was shown that the helicity of quiet sun flux obeys the hemispheric
rule. During their observations many ARs were present on the Sun and the authors
suggest that helicity pattern of the ARs could be reflected in the quiet sun
elements simply due to the origin of quiet sun flux from decayed flux of the
ARs.

%TBD ending here

\section{Discussion and Conclusions}
In this work we present for the first time synoptic (Carrington) maps of the
observed vector magnetic field.
These maps provide a representation of magnetic fields without any restrictive
assumptions about topology of fields, and therefore, are expected to improve the
outcome of coronal field extrapolation models using synoptic maps as input. Comparison of radial components
derived by traditional method using LOS magnetograms with the one derived from vector
data shows systematic differences in high and low-latitudes as well as in active regions.  Further, we show that the non vertical nature of magnetic field leads to systematic errors in the  radial field deduced from LOS observations, $B_{r(LOS)}$. The relative error, $|\Delta B_r| /|B_r|$ (\%) varies as tangent of the latitude and therefore becomes significant at high latitudes, even if inclination angle with respect to vertical direction remains same.
%The measurements near the poles pose severe difficulty due to obliqueness, however, such measurements are very important. Here we can emphasize that one must take advantage of large aperture of upcoming telescopes, like Advanced Technology Solar Telescope (ATST) \citep{keil2004}, and annual solar B-angle variation, as demonstrated by \cite{tsuneta2008}, to record vector fields near the poles. Presently SOLIS/VSM measurements are limited from Equator to $\sim$70$^\circ$ latitude and the results suggest that in this latitude range there is systematic tilt of the field from vertical direction.  Measurements very close to poles are expected to yield nearly vertical fields, as  the white light coronagraph images as well as global potential field models suggest.
The severity of the differences when using observed radial field as compared estimated radial field from LOS measurements, on the extrapolated coronal fields and solar wind derivations is presently unknown, and will be a subject of a separate future study.

%Further, these vector synoptic maps allow us to compute the
%distribution of vertical component of electric current density directly over the entire solar disk.
%The distribution of electric current density and its helicity provides a
%measure of non-potentiality of the solar magnetic fields, and it can be useful for
%understanding of solar dynamo, cycle and in space weather applications.
%Further, these maps could be used to carry out full sun extrapolations
%of the nonlinear force-free field (NLFFF) for global corona. Such
%extrapolations may provide essential information on global magnetic
%free energy budget of the solar corona for example, by taking out the potential
%energy from PFSS model
%extrapolations  based on true radial field. Also, such models may give an
%improved estimate of coronal holes and solar wind speed. We plan addressing some of these applications in our follow-up work.

As the first use for these new synoptic maps, we employed them
to study the distribution of current helicity density on the Sun.
We found that the hemispheric
pattern of current helicity density is present during ascending phase of cycle
24. Although the derived helicity maps do show patterns of opposite helicity present in both hemisphere,
there appears to be no indication of a systematic reversal in helicity at the beginning of Cycle 24
as predicted by some previous studies \citep[e.g.,][]{arnab2004}.
Neither we see a presence of well-defined bands of opposite helicity co-spatial with pattern of
torsional oscillations as was reported for solar cycle 22 \citep{pevtsov2003}.

%Further, we confirmed using direct observations the
%existence of hemispheric pattern of $H_c$ in high latitudes representative of
%large scale magnetic field. Such pattern was earlier deduced  from psuedo-
%vector reconstruction technique which uses longitudinal magnetograms
%\citep{pevtsov2000}.

We studied the hemispheric pattern for weak and strong fields separately
following criterion by \cite{mzhang2006}. Our results do show opposite sign of the hemispheric
preference for weak and strong fields. However, opposite to \cite{mzhang2006} we find that helicity
of strong fields follow the hemispheric rule, while helicity of weak fields exhibit inverse helicity sign-hemisphere
relation. Thus, for strong fields, the product of latitude ($\theta$) and current helicity (H$_c$) is negative
in agreement with the hemispheric rule (i.e., $\theta \cdot$H$_c <$ 0). For weak fields, $\theta \cdot$H$_c >$ 0.
Reasons for such disagreement between our results and those of \cite{mzhang2006} are unknown and need further investigation.

Despite these differences the important question that remains is: why
different hemispheric behavior of weak and strong fields is seen? The models
of helicity generation in solar magnetic fields must be confronted with
observational results. Mean field dynamo models based on $\alpha$-effect
\citep{steenbeck1966,seehafer1996} show two helicities, the one in the mean
field and another in the fluctuations. Further, both have similar magnitude
but opposite sign, such that the sign of helicity in the mean and fluctuating
fields is shown to be of positive and negative sign, respectively in Northern
hemisphere and vice versa in Southern. In the mean field dynamo the active
regions are thought of as fluctuations and not as the mean field. Thus  we get same hemispheric preference for sign of observed
current helicity in ARs  as we get in mean field dynamo models for
fluctuating field. What about the observational counterpart for helicity of
the mean field ? Could the mean latitudinal profile of helicity for weak
fields, which shows anti-hemispheric rule, as shown in right panel of
Figure~\ref{fig_strweak} represent the helicity of the mean field from dynamo
models, since they have same sign preference?

Another plausible explanation could
be that the current helicity observed in strong and weak magnetic field are a result of two different processes. For example, helicity in strong fields of active regions could be created by $\Sigma$-effect \citep{longcope1998}, while helicity of weak fields reflects their interaction with near surface flows. The twist can be induced in the magnetic flux due to vorticity of the (near) surface flows
\citep[e.g., supergranular flow, see][]{duvall2000}. The (near) surface flows
(from surface to a depth of about 16 Mm) were studied using ring-diagram technique by \cite{komm2007} as a
function of magnetic flux. They found that on average, quiet regions show
weakly divergent horizontal flows with small anticyclonic vorticity
(clockwise in the Northern hemisphere), while locations of high activity show
convergent horizontal flows with cyclonic vorticity (counterclockwise in
Northern hemisphere. Consider a untwisted flux tube embedded vertically in
photospheric layers and subject to clockwise horizontal flows. It can be seen
that a clockwise flow will induce positive twist in magnetic flux tube and a
counterclockwise flow will induce negative twist. Thus the flow patterns from
local helioseismology \citep{komm2007} would tend to induce positive current
helicity in the magnetic flux rooted in quiet regions and a negative current
helicity in the magnetic flux rooted in strong field regions, in the Northern
hemisphere. Such a  pattern is similar to what we see in our present observations, namely a positive helicity in weak field regions and negative helicity in strong field regions in the Northern hemisphere.
Thus, the observed behavior of current helicity in strong and weak fields could be explained in the framework of mean field dynamo models, or alternatively due to the interaction of magnetic flux tubes with turbulent plasma flows in the convection zone.

%In another study  of two uniformly twisted sunspots with chromospheric whirls around them \citep{komm2013}, one in Northern and another in Southern hemisphere, both of which followed hemispheric rule, it was found that the kinetic helicity of the subsurface flows derived using local helioseismology  was anti-correlated with the twist pattern seen in photospheric vector magnetogram of these two sunspots.

However,  more observations are needed to improve the statistics further.  Synoptic maps of vector magnetic field presented here appear to be useful tools for such statistical studies. Accumulation of more vector synoptic maps during the entire solar cycle 24 and beyond would be useful to establish patterns of helicity on the sun and will also help in testing the validity of various  physical mechanism(s) that have been proposed in order to explain the hemispheric preference of magnetic twist, its associated statistical dispersion and variation with solar cycle.

\acknowledgments
We thank the anonymous referee for his/her critical comments on the manuscript. We thank Lorraine Callahan for reducing/preparing data for inversions. This work utilizes SOLIS data obtained by the NSO Integrated Synoptic
Program (NISP), managed by the National Solar Observatory, which is
operated by the Association of Universities for Research in Astronomy (AURA),
Inc. under a cooperative agreement with the National Science Foundation.
Work by SG and AAP  was partially supported by NSF/SHINE Award No. 1062054 to the National Solar Observatory. GVR and SAA acknowledge support by the Ministry of Education and Science
of Russian Federation (GS 8407 and GK 14.518.11.7047) and  by the
RFBR (12-02-31746 mol\_a, 12-02-33110    mol\_a\_ved).

{\it Facilities:} \facility{SOLIS (VSM)}.

%\bibliographystyle{apj}
%\bibliography{apj-jour,helicity_ref}

\clearpage

\appendix
\section{Effect of Noise on $\Delta B_r/|B_r|$}
Let us assume that $B_r$  and $B_r(LOS)$ (or simply $B_L$) have same sign (say positive), and noise ($\sigma$) is small compared to magnitude of radial field (both $B_r$ and $B_L$)

Then,
$$\frac{\Delta B_r}{|B_r|}=\frac{|B_r\pm\sigma| - |B_L\pm\sigma|}{|B_r\pm\sigma|}= \frac{(B_r\pm\sigma) - (B_L\pm\sigma)}{B_r\pm\sigma} =\frac{(B_r -B_L) \pm \sigma}{B_r\pm\sigma}$$

If $\sigma << B_r$ and $B_r \approx B_L$, we can re-write the above as

$$\frac{(B_r-B_L)\pm\sigma}{B_r\pm\sigma}\approx\frac{\pm\sigma}{B_r}$$

\begin{figure}
\epsscale{0.6}
\plotone{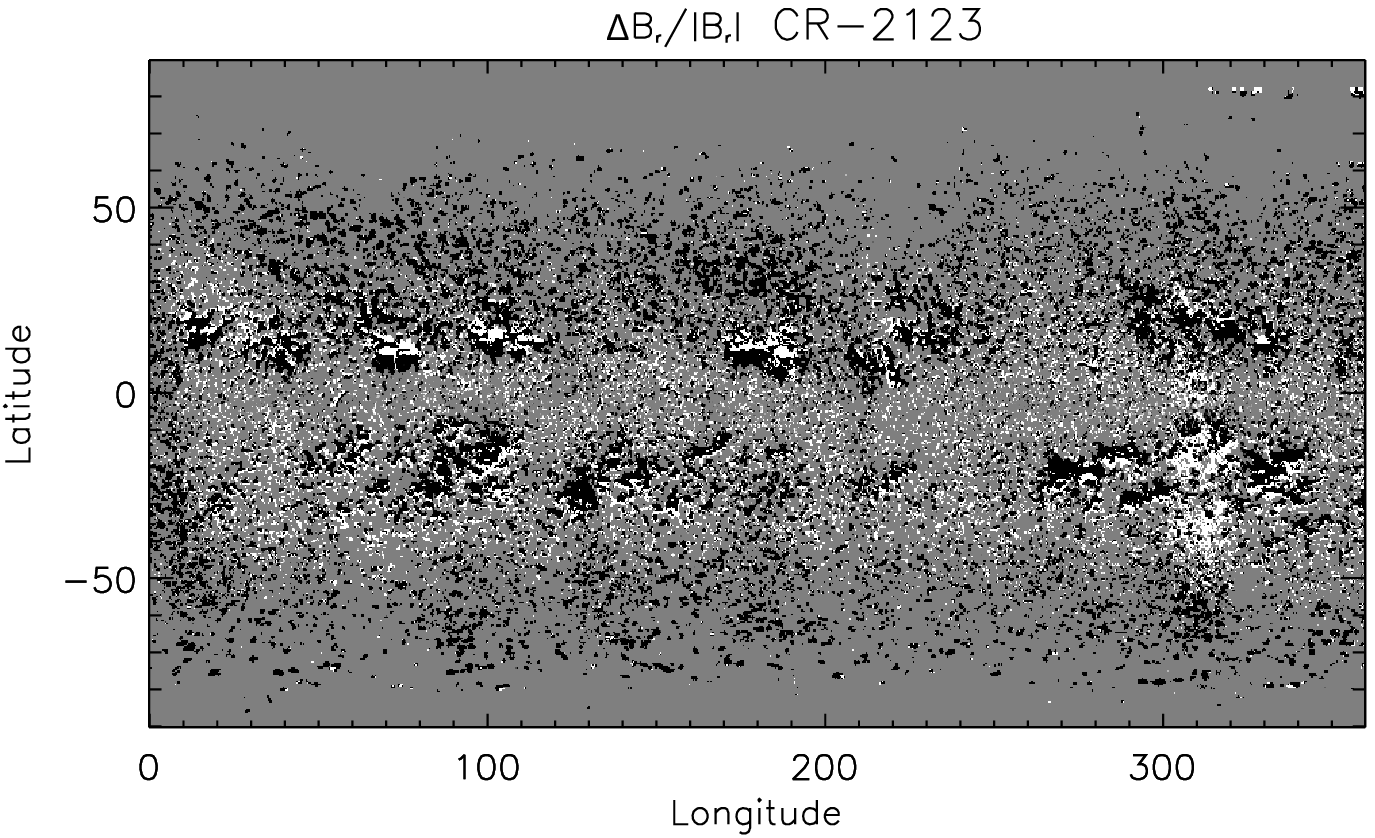}
\plotone{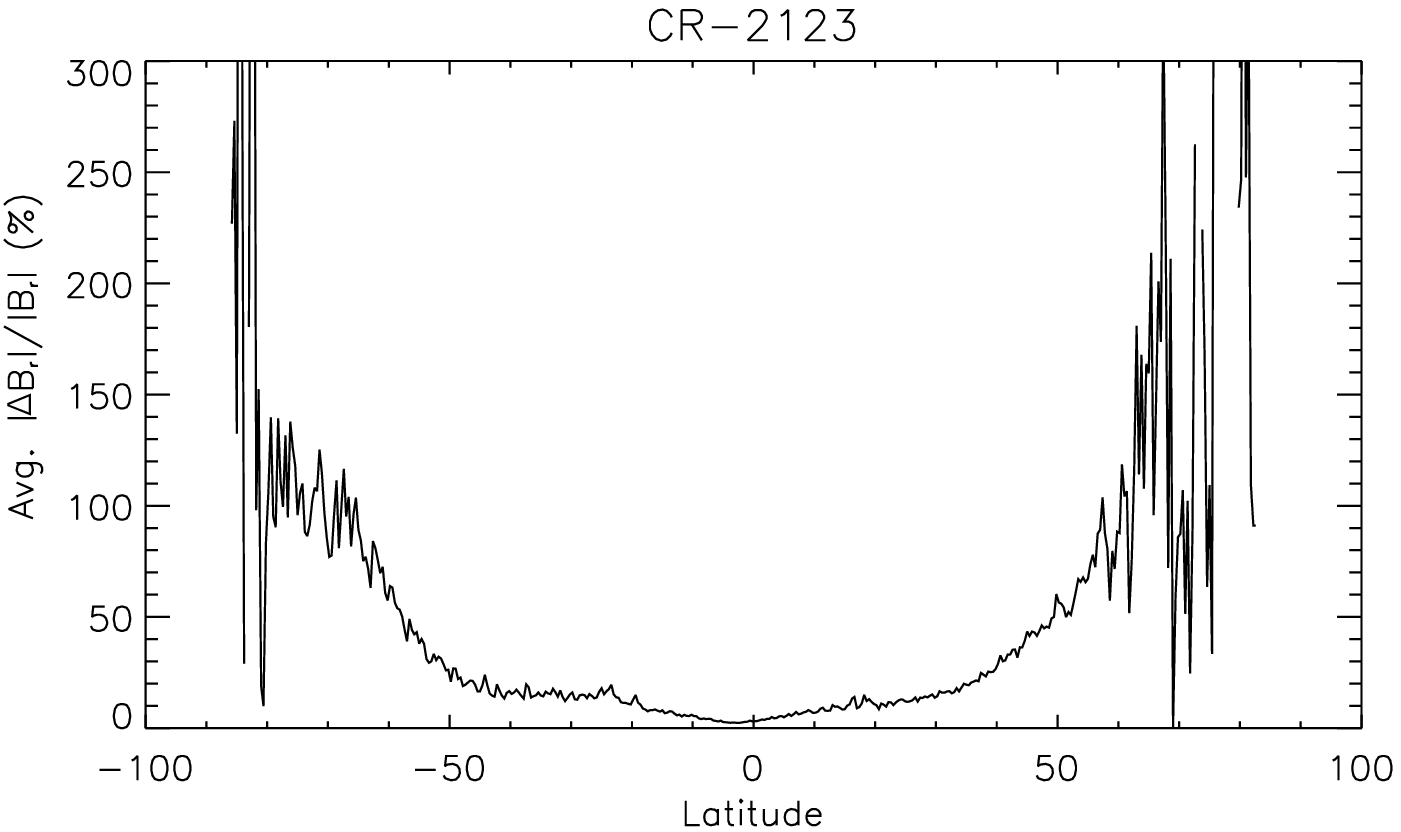}
\plotone{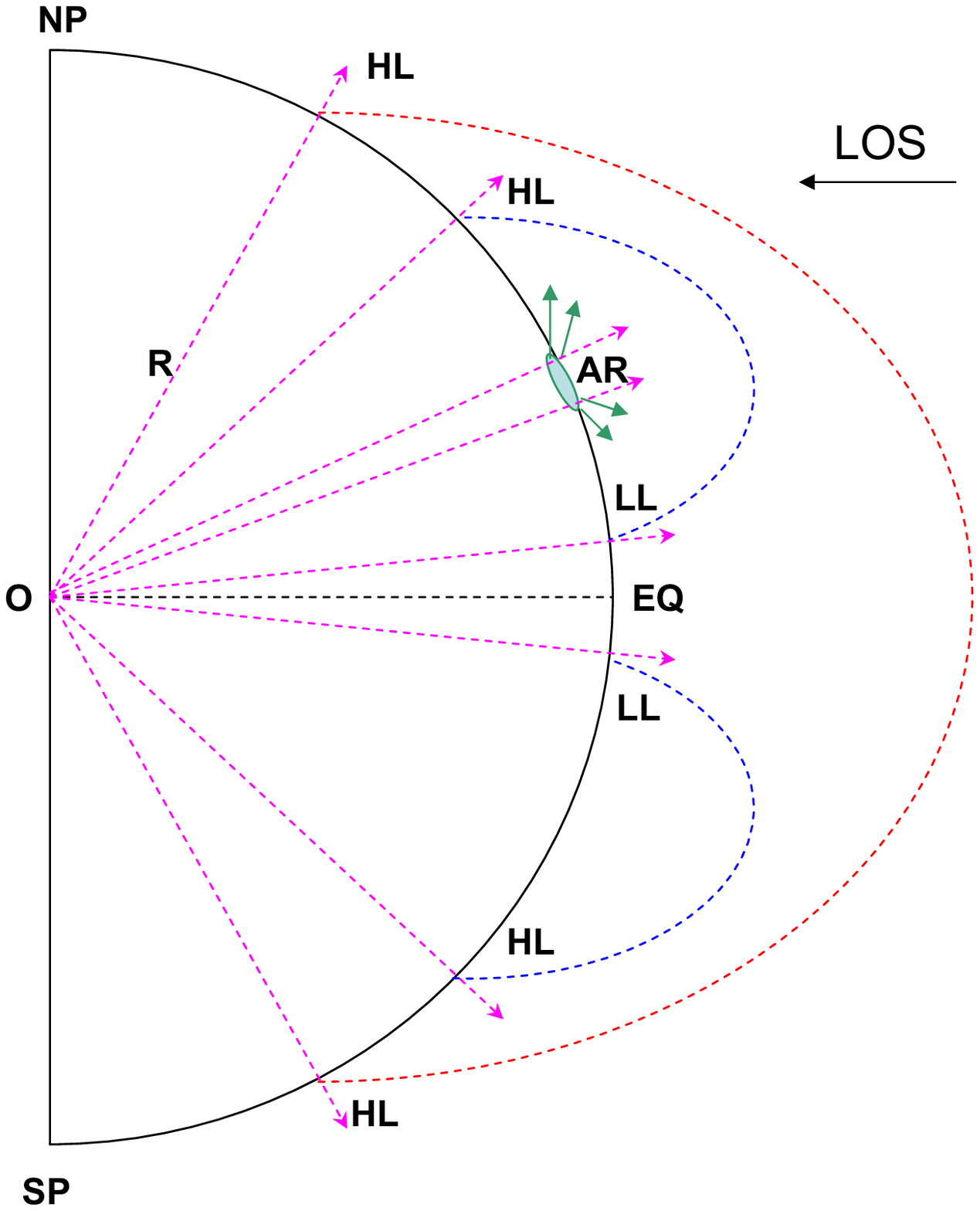}
\caption{Top panel displays the signed relative difference, $\Delta B_r/|B_r|$ (\%), between the absolute values of observed radial field, $|B_r|$, and the radial field derived from LOS magnetic field, $|B_r(LOS)|$, during Carrington rotation 2123.  The signed relative difference is  saturated between $\pm$5\% to emphasize the sign of the difference. The middle panel shows the longitudinal average of absolute relative difference, $|\Delta B_r|/|B_r|$ (\%). In the bottom panel we show a cartoon model to illustrate the sign pattern observed in the top panel.  }
\label{diffmaps}
\end{figure}

Thus, as noise ($\sigma$) increases, the scatter in the relative difference, $\Delta B_r / |B_r|$, also increases proportionally. A systematic latitudinal pattern in the sign of relative difference, $\Delta B_r / |B_r|$, as seen in top panel of Figure~\ref{diffmaps} is not expected from random noise in $B_r$ and $B_L$.

\clearpage

\clearpage

\clearpage

%\begin{figure}
%\epsscale{0.8}
%\plotone{brdiff_2123.eps}
%\plotone{av_brdiff_2123.eps}
%\plotone{av_brdiff_percent2123.eps}
%\caption{The top panel displays, as an example, the difference between the true radial field, $B_r$, and the radial field approximated from LOS magnetic field, $B_{r_los}$, under the assumption of verticality of solar magnetic field, during Carrington rotation 2123.  The difference signal is scaled between $\pm$ 50 G. The bottom panel shows the longitudinal average of the difference signal. }
%\label{diffmaps}
%\end{figure}

\clearpage

\clearpage

\clearpage

\clearpage

\clearpage

\end{document}